# Identifying the Largest RoCoF and Its Implications

Licheng Wang, Luochen Xie, Gang Huang, *Senior Member*, *IEEE*, Changsen Feng

*Abstract*—The rate of change of frequency (RoCoF) is a critical factor in ensuring frequency security, particularly in power systems with low inertia. Currently, most RoCoF security constrained optimal inertia dispatch methods and inertia market mechanisms predominantly rely on the center of inertia (COI) model. This model, however, does not account for the disparities in post-contingency frequency dynamics across different regions of a power system. Specifically, regional buses can exhibit significantly larger RoCoFs than that predicted by the system's COI, particularly in systems characterized by unevenly distributed inertia. In this letter, a post-contingency nodal RoCoF model is established, and the maximal initial RoCoF is further proven to occur at generator buses equipped with inertia, rather than at inertia-less load buses. This finding facilitates the development of the optimal nodal inertia dispatch method and the nodal inertia market mechanism in a convex and concise form. Our argument is further verified by the simulation results of the South East Australia power system under various scenarios.

## I. INTRODUCTION

KEEPING a certain level of inertia is crucial for the frequency stability of power systems. In traditional synchronous generator (SG) dominated power systems, there is often an abundance of inertia. However, the transition towards renewable energy sources (RES) poses a significant challenge to this inertia adequacy [1], potentially leading to extremely large rates of change of frequency (RoCoF) during frequency events.

Most RoCoF security constrained optimal dispatch methods rely on the center of inertia (COI) model. These methods typically enforce constraints on total inertia requirements for optimal generator dispatch to ensure COI RoCoF security [2, 3]. Recent years have also seen the development of inertia market mechanisms based on the COI model, where a uniform price for inertia market clearing is determined through the Lagrange multiplier of the COI-based RoCoF security constraint [4-6].

However, the initial post-contingency RoCoF varies across different regions, particularly in power systems with uneven inertia distribution. This variance can result in the regional RoCoF exceeding its threshold, potentially triggering additional generator trips, even while the COI RoCoF remains within the acceptable limit. Furthermore, the COI-based frequency transient model fails to account for the differential contributions of inertia from various system areas, thus lacking the granularity needed to signal where inertia is most required.

To bridge this gap, this letter establishes a nodal RoCoF model and indicates for the first time that the largest nodal RoCoF following a disturbance will be at one of the generator buses. This finding provides a concise paradigm for nodal RoCoF security constraints in optimal inertia dispatch and offers system operators an approach to price inertia services based on both the volume and location of inertia.

## II. NODAL ROCOF MODEL

### A. Distribution of Power Impacts at $0^+$

A power disturbance occurs at time instant 0 will immediately change the power flow distribution across an electricity network. Correspondingly, variations of bus injection power and phase angles of an $n$-generator $m$-load system at $0^+$ (immediately after 0) can be expressed through an augmented direct current (DC) power flow model [7] as

$$\begin{bmatrix} \Delta P_G(0^+) \\ \Delta P_D(0^+) \end{bmatrix} = \begin{bmatrix} B_{GG} & B_{GB} \\ B_{BG} & B_{BB} \end{bmatrix} \begin{bmatrix} \Delta \Theta_G(0^+) \\ \Delta \Theta_D(0^+) \end{bmatrix}, \quad (1)$$

where $\Delta P_G \in \mathbb{R}^{n\times 1}$, $\Delta P_D \in \mathbb{R}^{m\times 1}$, $\Delta \Theta_G \in \mathbb{R}^{n\times 1}$, $\Delta \Theta_D \in \mathbb{R}^{m\times 1}$ are vectors of variations of generators' electromagnetic power, load demand, phase angles behind the internal generator reactance, phase angles of load buses, respectively; $B_{GG}$, $B_{GB}$ and $B_{BG}$ are susceptance matrices obtained using the internal reactance of the generators; $B_{BB}$ is the sum of a standard network susceptance matrix and a diagonal matrix that accounts for internal reactance of generators. In this letter, we assume: 1) all load buses are regarded as inertia free; 2) the internal transient control process of virtual SGs is ignored, and therefore inertia and virtual inertia are not discriminated.

Due to the intrinsic nature of inertia, SGs and virtual SGs will keep their rotor angles constant from time 0 to $0^+$. Therefore, all elements of $\Delta \Theta_G(0^+)$ are equal to 0, namely,

$$\Delta \Theta_G(0^+) = [0, \dots, 0]^T. \quad (2)$$

According to [8], load power consumption can be regarded as constant during the contingency. Therefore, power variations of load buses are equal to zero except for the one where the disturbance occurs. Correspondingly, we have

$$\Delta P_D(0^+) = [0, \dots, \Delta p_{D,j}(0^+), \dots, 0]^T \quad (3)$$

$$\Delta p_{D,j}(0^+) = P_{dis}, \ j \in \mathcal{D}, \quad (4)$$

where $\Delta p_{D,j}$ is the $j^{\text{th}}$ element of $\Delta P_D$; $P_{dis}$ represents the size of the power disturbance; $\mathcal{D}$ denotes the set of load buses. Both $\Delta \Theta_G(0^+)$ and $\Delta P_D(0^+)$ are known vectors as in (2) and (3), and substituting (2) and (3) into (1), we have the distribution of power impacts at different generator buses as

$$\Delta P_G(0^+) = B_{GB} B_{BB}^{-1} \Delta P_D(0^+). \quad (5)$$

This work was supported by the National Natural Science Foundation of China (52007170, U22B20116).

Licheng Wang (e-mail: wanglicheng@zjut.edu.cn), Luochen Xie, and Changsen Feng are with the College of Information Engineering, Zhejiang University of Technology, Hangzhou 310023, China.

Gang Huang is with Zhejiang University, Hangzhou 311121, China (corresponding author, e-mail: huanggang@zju.edu.cn).

***Proposition 1:*** *If the $n$-generator $m$-load power system is interconnected, the matrix $\boldsymbol{B_{BB}}$ is invertible.*

***Proof:*** An interconnected $n$-generator $m$-load power system can be divided into a passive system (with only load buses) and generator buses as shown in Fig. 1 (a). According to its definition, $\boldsymbol{B_{BB}} \in \mathbb{R}^{m \times m}$ is the sum of a standard network susceptance matrix $\boldsymbol{B_{BB}^0} \in \mathbb{R}^{m \times m}$ that ignores all shunt elements and a diagonal matrix $diag[\ldots, \sum_i b_{ji}, \ldots]$ as

$$\boldsymbol{B_{BB}} = \boldsymbol{B_{BB}^0} + diag[\ldots, \sum_i b_{ji}, \ldots], \; j \in \mathcal{D}, i \in \mathcal{G}, \quad (6)$$

where $b_{ji}$ is the susceptance between load bus $j$ and generator bus $i$, and $b_{ji} = 0$ if they are not directly connected; $\mathcal{G}$ denotes the set of generator buses. After all generator buses are grounded and share the same zero-potential point as in Fig. 1 (b), the dimension of the system susceptance matrix (denoted as $\boldsymbol{B'} \in \mathbb{R}^{m+1 \times m+1}$) will reduce from $m+n$ to $m+1$. Correspondingly, we have

$$\boldsymbol{B'} = \begin{bmatrix} \boldsymbol{B_{BB}} & \vdots \\ & -\sum_i b_{ji} \\ \ldots & -\sum_i b_{ji} & \ldots & \sum_j \sum_i b_{ji} \end{bmatrix}, \boldsymbol{B''} = \begin{bmatrix} \boldsymbol{B_{BB}} & \vdots \\ & -\sum_i b_{ji} \\ & \vdots \end{bmatrix}. (7)$$

It is worth noting, $\boldsymbol{B'}$ is also the Laplacian matrix of a connected undirected graph, and its rank is one order less than its dimension, namely $rank(\boldsymbol{B'}) = m$, according to the graph theory [10]. Furthermore, since each column of matrix $\boldsymbol{B'}$ sums to zero (considering the property of susceptance matrices without shunt elements), its last row can be expressed as a linear combination of its first $m$ rows. Therefore, the rank of its submatrix $\boldsymbol{B''} \in \mathbb{R}^{m \times m+1}$, which is constituted by the first $m$ rows of $\boldsymbol{B'}$, is $m$. Similarly, since each row of matrix $\boldsymbol{B''}$ sums to zero, the last column (from the left side to the right side) of $\boldsymbol{B''}$ can be expressed as a linear combination of its first $m$ columns. Considering both the linear combination property and $rank(\boldsymbol{B''}) = m$, we can finally conclude, columns of $\boldsymbol{B_{BB}} \in \mathbb{R}^{m \times m}$, that are also the first $m$ columns of $\boldsymbol{B''}$, are linearly independent, and $\boldsymbol{B_{BB}}$ is invertible. ∎

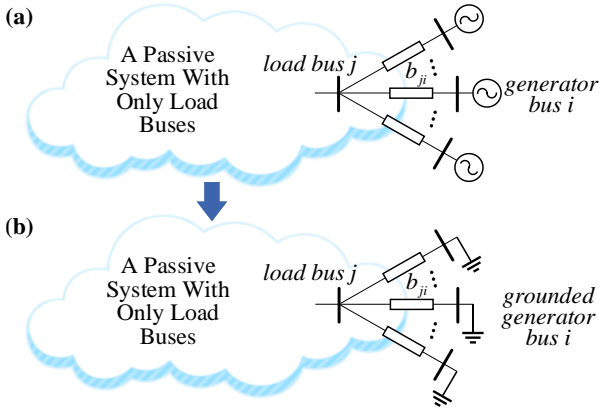

Fig. 1 (a) An interconnected power system, (b) An interconnected power system with grounded generator buses.

*B. Initial RoCoF at Different Buses*

The initial RoCoF after a disturbance is defined as the RoCoF at $0^+$. For generator bus $i$, its local RoCoF can be obtained through the swing equation:

$$RoCoF_{G,i}(0^+) = \frac{\Delta p_{G,i}(0^+)}{2H_i}, \; i \in \mathcal{G}, \quad (8)$$

where $RoCoF_{G,i}$ represents the local RoCoF of generator bus $i$; $\Delta p_{G,i}$ is the $i^{\text{th}}$ element of $\boldsymbol{\Delta P_G}$; $H_i$ represents the sum of synchronous inertia and virtual inertia on bus $i$.

Since $\boldsymbol{\Delta P_D}$ is time invariant after the disturbance, taking the derivative of (1) with respect to time for twice, we have

$$RoCoF_D(0^+) = \boldsymbol{T} \cdot RoCoF_G(0^+) \quad (9)$$
$$\boldsymbol{T} = -\boldsymbol{B_{BB}^{-1}} \boldsymbol{B_{BG}}, \quad (10)$$

where $RoCoF_G \in \mathbb{R}^{n \times 1}$ and $RoCoF_D \in \mathbb{R}^{m \times 1}$ are regional RoCoF vectors of generator and load buses, respectively; $\boldsymbol{T}$ is a known matrix with elements depending on the system topology as well as line (generator) parameters.

***Proposition 2:*** *For matrix $\boldsymbol{T}$ as defined in (10), if the power system is interconnected, each row of $\boldsymbol{T}$ sums to one.*

***Proof:*** Elements of $\boldsymbol{B_{BG}} \in \mathbb{R}^{m \times n}$, $\boldsymbol{B_{BB}} \in \mathbb{R}^{m \times m}$ and $\boldsymbol{T} \in \mathbb{R}^{m \times n}$ are defined as

$$\boldsymbol{B_{BG}} = [a_{ij}]_{j=1,\ldots,n}^{i=1,\ldots,m}, \boldsymbol{B_{BB}} = [b_{ij}]_{j=1,\ldots,m}^{i=1,\ldots,m}, \boldsymbol{T} = [c_{ij}]_{j=1,\ldots,n}^{i=1,\ldots,m}. (11)$$

Considering the property of a susceptance matrix without shunt elements, the sum of elements in each row is equal to zero. Specifically, for the susceptance matrix in (1), we have

$$-\sum_{j=1}^{n} a_{ij} = \sum_{j=1}^{m} b_{ij}, \; \forall i. \quad (12)$$

Equation (10) can be equivalently rewritten as

$$\boldsymbol{B_{BB}} \boldsymbol{T} = -\boldsymbol{B_{BG}}. \quad (13)$$

Substituting (11) into (13) and calculating the sum of each row's elements, we have

$$\sum_{j=1}^{n} \sum_{k=1}^{m} b_{ik} c_{kj} = -\sum_{j=1}^{n} a_{ij}, \; \forall i. \quad (14)$$

Substituting the right side of (14) by (12), we get

$$\sum_{j=1}^{n} \sum_{k=1}^{m} b_{ik} c_{kj} = \sum_{j=1}^{m} b_{ij}, \; \forall i, \quad (15)$$

or equivalently

$$\sum_{k=1}^{m} [b_{ik} \sum_{j=1}^{n} c_{kj}] = \sum_{k=1}^{m} b_{ik}, \; \forall i. \quad (16)$$

Subtracting the left side of (16) from its right side, we have

$$\sum_{k=1}^{m} [b_{ik}(1 - \sum_{j=1}^{n} c_{kj})] = 0, \; \forall i, \quad (17)$$

namely

$$\boldsymbol{B_{BB}}[\ldots, \; (1 - \sum_{j=1}^{n} c_{kj}), \; \ldots]^T = \boldsymbol{0}, \quad (18)$$

where $[\ldots, \; (1 - \sum_{j=1}^{n} c_{kj}), \; \ldots]^T \in \mathbb{R}^{m \times 1}$ is a column vector. Since $\boldsymbol{B_{BB}}$ is invertible as proven in ***Proposition 1***, (18) indicates $1 - \sum_{j=1}^{n} c_{kj} = 0$ holds for any row $k$. In other words, each row of matrix $\boldsymbol{T}$ sums to one. ∎

According to (9), the regional RoCoF of each load bus is a linear combination of that of generator buses. Further considering ***Proposition 2***, we can finally conclude that ***the largest RoCoF will be at one of the generator buses in an interconnected power system***.

## III. APPLICATIONS

The identification of the largest RoCoF and the nodal RoCoF model developed in Section II can be applied in the formulation of the optimal inertia dispatch problem, as shown in (19), with RoCoF security constraints only imposed on generator buses.

$$Min \ \sum_i c_i(H_i^v), \ i \in \mathcal{G} \quad (19a)$$
$$-2RoCoF_{max}(H^0 + H^v) \leq B_{GB}B_{BB}^{-1}\Delta P_D(0^+) : \underline{\sigma} \quad (19b)$$
$$B_{GB}B_{BB}^{-1}\Delta P_D(0^+) \leq 2RoCoF_{max}(H^0 + H^v) : \overline{\sigma} \quad (19c)$$
$$H^0 \leq H^0 + H^v \leq H^{max}, \quad (19d)$$

where $c_i(\cdot)$ in (19a) is the cost function of virtual inertia at generator bus $i$; (19b) and (19c), derived from (8)~(10), are regional RoCoF security constraints for generator buses, and $\underline{\sigma}, \overline{\sigma} \in \mathbb{R}^{n \times 1}$ are corresponding dual variables; $H^0, H^v, H^{max} \in \mathbb{R}^{n \times 1}$ are column vectors of synchronous inertia $H_i^0$, virtual inertia $H_i^v$ and the upper limits, respectively; (19d) is the upper and lower bounds of total inertia at generator buses. Notably, $H_i^0$ are known values in (19) according to the day ahead unit commitment scheduling, and $H_i^v$ are the decision variables that will only be applied to generator buses as the supplement of synchronous inertia for system frequency support.

Furthermore, similar to the locational marginal pricing (LMP) method, the shadow price $\rho_i$ of nodal virtual inertia at generator bus $i$ can be expressed as

$$\rho_i = 2RoCoF_{max}\left(\overline{\sigma_i} + \underline{\sigma_i}\right), \quad (20)$$

where $\underline{\sigma_i}$ and $\overline{\sigma_i}$ are the $i^{th}$ element of $\underline{\sigma}$ and $\overline{\sigma}$, respectively.

## IV. CASE STUDY

A five-area interconnected power system in South East Australia [9] with its fill dynamic is simulated in PSS®E as case studies in this letter. Fig. 2 validates the conclusion we made in Section II that the largest RoCoF after a contingency will occur at a generator bus under different scenarios. Specifically, Fig. 2 (a) and (c) are Scenarios 1 and 3, where the total inertia of the system is 53.25GWs and a sudden power disturbance of 150MW occurs at Bus 405 and Bus 306, respectively. The first digit of the bus number indicates the area it belongs to. In these two scenarios, the largest initial RoCoF are -1.206Hz/s at Bus 404 and -1.279Hz/s at Bus 302, respectively. Fig. 2 (b) corresponds to Scenario 2 where the system total inertia is reduced to 34.05GWs due to RES generation replacement. In this situation, a sudden load increase of 150MW at Bus 405 will induce the largest RoCoF of -1.834Hz/s on Bus 404 at time instant $0^+$. As indicated in Fig. 2, both Bus 404 and Bus 302 that have largest initial RoCoF under different scenarios are generator buses.

In order to keep the post-contingency initial RoCoF of all buses across the power system within an allowable range (e.g., -1Hz/s~1Hz/s in this letter), proper virtual inertia support is required. For example, extra virtual inertia supports of 58.4MWs at Bus 402 and 132.1MWs at Bus 404 are added as the supplement of synchronous inertia in the intra-day dispatch in Scenario 2. As a result, after a power disturbance of 150MW occurs at Bus 405, all initial nodal RoCoF across the power system will be within the safe range. As shown in Table I, the corresponding shadow prices of virtual inertia at Bus 402 and Bus 404 are 74.5$/MWs and 123.5$/MWs, respectively, according to the nodal inertia market clearing model as in (19)~(20).

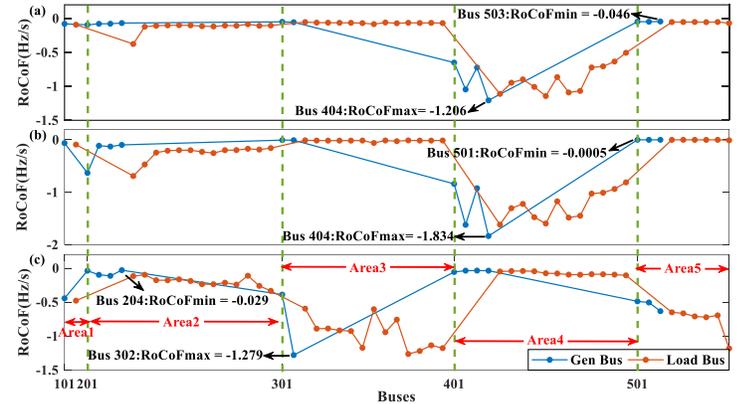

Fig. 2 Initial nodal RoCoF under different scenarios.

TABLE I
VIRTUAL INERTIA VOLUME AND SHADOW PRICES ON DIFFERENT BUSES

|  | Bus 402 | Bus 404 |
| --- | --- | --- |
| Virtual Inertia Volume | 58.4MWs | 132.1MWs |
| Shadow Prices | 74.5$/MWs | 123.5$/MWs |

## V. CONCLUSION

The post-contingency nodal RoCoF model has been established in this letter, and the maximum initial RoCoF is proven to occur at one of the generator buses. As a result, RoCoF security constraints only need to be applied to generator buses, which helps to establish a convex optimal nodal inertia dispatch scheme and a nodal inertia market clearing model. Simulation results of the South East Australian power system further verify the effectiveness of the largest RoCoF identification under different operation scenarios, as well as the proposed nodal inertia dispatch method and the nodal inertia pricing mechanism.